\documentclass[aps,prl,showpacs,groupedaddress]{revtex4}

\usepackage[dvips]{graphicx}
\usepackage{dcolumn}

\begin{document}

\title
{Vortex tilt modulus in Fulde-Ferrell-Larkin-Ovchinnikov state
}

\author{Ryusuke Ikeda}

\affiliation{%
Department of Physics, Kyoto University, Kyoto 606-8502, Japan
}

\date{\today}

\begin{abstract} 
Vortex tilt response in a Fulde-Ferrell-Larkin-Ovchinnikov (FFLO) vortex lattice is studied as a probe reflecting the spatial structure of this state. In quasi 2D materials under a parallel magnetic field, the tilt modulus $E_{2}$ {\it of the nodal planes} in the FFLO state modulating along the field decreases as the paramagnetic effect is effectively enhanced, and this reduction of $E_{2}$ in turn reduces the vortex tilt modulus. This reduction, more remarkable in higher fields or in more 2D-like systems, of vortex tilt modulus upon entering the FFLO state may be one origin of an anomalous anisotropic reduction of sound velocity detected in an ultrasound measurement in CeCoIn5. 
\end{abstract}

\pacs{74.25.Dw, 74.25.Ld, 74.70.Tx, 74.81.-g}

%\keyword{}

\maketitle

\section{I. Introduction} 

Recent accumulating experimental facts in heat capacity \cite{Bianchi}, thermal conductivity \cite{Capan}, penetration depth \cite{Martin}, and NMR data \cite{Kaku} certainly indicate the presence of a new superconducting (SC) phase of CeCoIn5 at low $T$ and under high magnetic fields {\it parallel} to the SC layers. This new phase, separated from the ordinary Abrikosov vortex lattice via a second order transition, is expected to be the FFLO vortex lattice with one-dimensional periodic modulation of the SC order parameter $\Delta$ {\it parallel} to the field ${\bf H}$ based on a consistency between these observations and a recent theory \cite{AI,IA2} on the characters of transitions between different phases. If the FFLO modulation is {\it perpendicular} to ${\bf H}$, the mean field $H_{c2}$-transition between the normal and the FFLO state is usually expected to be of second order \cite{RIM2S}, just like the conventional result in the Pauli limit \cite{Pauli}. However, this is incompatible with the discontinuous $H_{c2}$-transition \cite{Izawa,Bianchi} in CeCoIn5. Spatial structures of a FFLO state may also be reflected in its elastic properties, and the tilt response of vortices should be sensitive to the direction of the periodic modulation. 

%%%%%%%%%%%%%%%%%%%
\begin{figure}[t]
\scalebox{1.1}[1.1]{\includegraphics{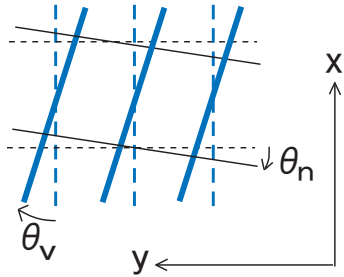}}
\caption{Sketch representing tilted vortex lines (thick solid lines) and nodal planes (thin solid ones) in the FFLO state modulating along ${\bf H}$ of a quasi 2D system in a parallel field (${\bf H} \parallel {\hat x}$). Thick and thin dashed lines denote their positions in equilibrium. The SC layer and the anisotropy axis correspond to the $x$-$y$ plane and the $z$-axis, respectively. } \label{fig.1}
\end{figure}
%%%%%%%%%%%%%%%%%

In this paper, we examine changes of vortex tilt modulus occurring through the transition between the FFLO vortex lattice with modulation parallel to ${\bf H}$ and the familiar Abrikosov lattice and show that, through a coupling between the vortices and the nodal planes accompanying the FFLO modulation , a measurable reduction of vortex tilt modulus may occur in such an FFLO state of uniaxially anisotropic superconductors in fields parallel to the SC layers. The present result may be relevant to the ultrasound experiment in CeCoIn5 \cite{ISSP}, in which a {\it monotonous} reduction of sound velocity upon cooling through $H_{\rm FFLO}(T)$ was observed only for a sound mode accompanied by vortex tilts. 

First, a qualitatively expected feature of the tilt response in the FFLO state will be explained. The FFLO state of our interest is the so-called LO state in which $\Delta$ in equilibrium has a periodic modulation with a period $2 \pi/Q$ parallel to ${\bf H} \parallel {\hat x}$ and vanishes on periodic nodal planes lying in $y$-$z$ plane (see Fig.1). The continuous FFLO to Abrikosov transition at $H_{\rm FFLO}(T)$ is characterized by a vanishing of the $Q^2$ term in the mean field free energy. On the other hand, the vortex line tension $\delta C_{44}$ of the Abrikosov lattice, which is one part of its total tilt modulus 
\begin{equation}
C_{44} = \delta C_{44} + \frac{B^2}{4 \pi}, 
\end{equation}
is defined from the gradient term in the {\it fluctuation} free energy for variations parallel to ${\bf H}$, where $B$ is the uniform flux density. Then, as in the conventional elastic softening in the ordinary solids, it is natural to expect \cite{AI} a tilt softening to occur on $H_{\rm FFLO}(T)$ with a cusp-like minimum of $\delta C_{44}$ (see Fig.2). Although such a behavior is generally expected for the so-called FF state with no nodal planes of the amplitude $|\Delta|$ and presumably also for other modulated vortex lattices with nodal planes parallel to the vortices, this picture is, when applied to the LO state with periodic nodal planes perpendicular to the vortices, justified only in the limited case where the nodal planes are never coupled with the vortices. In the LO state, a tilted nodal plane can carry the magnetic flux, and hence, a vortex tilt is induced by a small tilt of nodal planes since the number of vortices should remain unchanged for a {\it small} variation. This statement is represented in terms of tilt angles $\theta_v$ and $\theta_n$ of the vortices and of the nodal planes (see Fig.1), respectively, by the elastic energy 
\begin{equation}
\frac{\delta {\cal F}(u,s)}{N(0) T_c^2} = \biggl\langle \frac{1}{2} E_{1} \theta_v^2 + \frac{1}{2} E_{2} \theta_n^2 - 2 E_{3} \theta_n \, \theta_v \biggr\rangle, 
\label{elasF}
\end{equation}
where $N(0)$ is the density of states per spin in the normal state, and the bracket $\langle \, \, \, \rangle$ denotes the spatial average. The dimensionless vortex line tension defined under fixed nodal planes is given by $E_{1}$, $E_{2}$ is the corresponding tilt modulus of the nodal 
planes, and a nonvanishing coefficient $E_{3}$ of the coupling term is a consequence of the periodic modulation of the equilibrium $\Delta$. Hereafter, the nonlocality arising from the long interaction range between the vortices is neglected in $C_{44}$ \cite{com2}. Then, the vortex line tension $\delta C_{44}$ in the FFLO state stabilized by a positive $E_{2}$ becomes 
\begin{equation}
\delta C_{44} = N(0) T_c^2 \biggl[ E_{1} - 4 \frac{(E_{3})^2}{E_{2}} \biggr].
\label{delc44}
\end{equation}
Equation (\ref{delc44}) implies that $\delta C_{44}$ is reduced more drastically with decreasing $E_{2}$. Such a decrease of $E_{2}$ occurs due to an effective reduction of the orbital depairing because the modulation parallel to ${\bf H}$ is supported by the orbital effect of the magnetic field. In quasi 2D superconductors under a field parallel to the SC layers, the orbital depairing effect becomes less important in more 2D-like systems where the paramagnetic depairing is relatively more important. In the Pauli limit with no orbital depairing, the direction of modulation is {\it spontaneously} chosen, i.e., $E_2$ is vanishingly small, as far as a Fermi surface (FS) anisotropy is negligible. Thus, in highly 2D-like systems and/or a case with a larger Maki parameter $\alpha_{\rm M}$, the FFLO state in ${\bf H} \parallel {\hat x}$ should show a softer tilt response, for displacements $\parallel {\hat y}$, as a consequence of a large fluctuation of nodal planes induced by an enhanced paramagnetic depairing. 
%%%%%%%%%%%%%%%%%%%
\begin{figure}[b]
\scalebox{0.25}[0.25]{\includegraphics{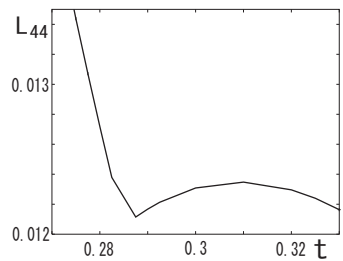}}
\caption{Example of $t$ ($=T/T_c$) dependence of $L_{44}$, proportional to $\delta C_{44}$, in ${\bf H} \parallel {\hat z}$ which has a cusp-like minimum at the FFLO to Abrikosov transition temperature $0.288 T_c$.} 
\label{fig.2}
\end{figure}
%%%%%%%%%%%%%%%%%

The physical argument given above implies that, in extremely 2D-like systems in the parallel fields, $\delta C_{44}$ in the FFLO state may take a {\it negative} value and suggests a possibility that even a tilt instability of the FFLO state might occur. To see to what degree the reduction of $C_{44}$ is substantial in real systems, a consistent and microscopic derivation of $E_n$ ($n=1$, $2$, and $3$) and the phase diagram will be performed in the remainder of this paper, and the ultrasound data \cite{ISSP} will be discussed based on the resulting tilt response. 

This paper is organized as follows. In sec.II, details of the model and our calculations performed to obtain the phase diagram and the tilt modulus are explained. In sec.III, typical examples of our numerical results following from the expressions derived in sec.II and Appendices are shown. In Appendix A, expressions following from a higher LL mode leading to a correction to $H_{c2}$ are given, and a theoretical background on the relation between the vortex tilt deformation and a LL mode of the order parameter is explained in details in Appendix B. 

\section{II. Model and Description}

 Our approach for describing the vortex lattices affected by the paramagnetic depairing takes the same route as the previous one \cite{AI} and starts with a BCS Hamiltonian ${\cal H} = {\cal H}_0 + {\cal H}_J + {\cal H}_{\rm int}$ for quasi 2D systems, where 
\begin{eqnarray}
{\cal H}_0 &=& d \sum_{\sigma, j}  \int d^2 r_{\perp} 
        ( \, {\varphi}_{j}^{\sigma}({\bf r_\perp}) \, )^\dagger \Bigg[ \, 
        \frac{ ({\rm -i}{\nabla}_{\perp} + e{\bf A} )^2}{2m} - \sigma \mu H \, 
        \Bigg] {\varphi}_{j}^{\sigma}({\bf r_\perp}), 
\label{BCSkin}
\end{eqnarray} 

%%%%%
\begin{eqnarray}
{\cal H}_J &=& - \frac{J \, d}{2} \sum_{\sigma, j}  \int d^2 r_{\perp} \Bigg(
{\varphi}_{j}^{\sigma \dagger}({\bf r_\perp})
{\varphi}_{j+1}^{\sigma}({\bf r_\perp}) +  {\varphi}_{j+1}^{\sigma \dagger}({\bf r_\perp}) {\varphi}_{j}^{\sigma}({\bf r_\perp}) \Bigg),
\label{interlayer}
\end{eqnarray}
%%%%%
and 
\begin{eqnarray} 
{\cal H}_{\rm int} &=& - \frac{|g| \, d}{2} \sum_{\sigma, j} 
 \int \frac{d^2 k_{\perp}}{(2 \pi)^2} 
	B_{\sigma, j}^{\dagger}({\bf k_\perp})
          B_{\sigma, j}({\bf k_{\perp}})  
\label{BCSint}
\end{eqnarray}
with 
$B_{\sigma, j}({\bf k_{\perp}}) = \sum_{\bf p_\perp} {\hat \Delta}_{\bf p}
a_j^{-\sigma}({\bf -p_-}) a_j^{\sigma}({\bf p_+})$, where ${\bf p}_\pm = {\bf p_\perp \pm {\bf k}_\perp}/2$. Here, $j$ is the index numbering the SC layers, ${\bf p}_\perp$ is the component of ${\bf p}$ parallel to the layers, ${\hat \Delta}_{\bf p}$ is the normalized orbital part of the pairing-function which, in the case of $d_{x^2-y^2}$-pairing, is written as 
$\sqrt{2}({\hat p}_x^2 - {\hat p}_y^2)$ in terms of the unit vector ${\hat {\bf p}}$ parallel to the layers, and $m$ is the effective mass of a quasi-particle. Further, $\sigma \mu B = \mu B$ or $-\mu B$ is the Zeeman energy \cite{comhb}. In discussing our calculation results, the strength of the paramagnetic effect will be measured by the Maki parameter $\alpha_M = \sqrt{2} H_{\rm orb}(0)/H_P(0)$, 
where $H_{\rm orb}(0)$ and $H_P(0) = \pi T_c/(\sqrt{2} e^{\gamma_{\rm E}} \, \mu) \simeq 1.2 T_c/\mu$ are the orbital and Pauli limiting fields at $T=0$ defined within the weak-coupling BCS model, 
respectively, where $\gamma_{\rm E}=0.577$ is an Euler constant. 
Hereafter, the gauge field ${\bf A}$ will be assumed to consist only of the contribution of the uniform flux density $B$, i.e., we work in the type II limit with no spatial variation of flux density, because we are interested mainly in the field region near $H_{c2}$. 

We use the familiar quasi-classical approximation for the single-particle propagator
%%%%%
\begin{equation}
G_{\varepsilon, \sigma}^B({\bf r,r'}) 
= G_{\varepsilon, \sigma}({\bf r-r'}) 
  e^{{\rm i} e\int_{\bf r}^{\bf r'} d{\bf s \cdot A} }.
\label{Green1}
\end{equation}
%%%%%
Here, the Fourier transform of $G_{\varepsilon, \sigma}({\bf r})$ 
is given by 
%%%%%
\begin{equation}
G_{\varepsilon, \sigma}({\bf p})
=\Big[{\rm i}{\varepsilon} + \sigma \mu B - \varepsilon_{\bf p} \Big]^{-1} ,
\label{Green2}
\end{equation}
%%%%%
where $\varepsilon_{\bf p} = ({\bf p}_\perp^2 - p_F^2)/(2 m) - J\cos(p_z s)$, $p_F$ is the Fermi momentum in 2D case, and 
$\varepsilon_n$ denotes the Matsubara frequency 
$2 \pi T (n+1/2)$. 
Since we take account of the paramagnetic depairing suppressing the upper critical field {\it in the mean field approximation} $H_{c2}(T)$, 
the use of the quasi-classical approximation, valid if $p_F r_B \gg 1$, 
is safely valid, where $r_B=(2|e| B)^{-1/2}$ is the magnetic length. 

Hereafter, in deriving an appropriate Ginzburg-Landau (GL) functional in ${\bf H} \parallel {\hat x}$, the spatial variation of the SC order parameter field $\Delta$ in the out-of-plane ($z$-) direction is assumed to have a longer range than the interlayer spacing $d$. When the paramagnetic effect is absent, this continuum approximation is valid only if $B \ll B_{\rm cr} \equiv 1/(2e \, \gamma \, d^2)$ \cite{II}. Here, $\gamma$ ($> 1$) is the anisotropy defined from the ratio between the in-plane coherence length $\xi_0 = v_F/(2 \pi T_c)$ and the out-of-plane coherence length $\xi_{0c}$, where $v_F$ is the Fermi velocity in 2D case. In the present case where the $H_{c2}(T)$ is reduced via the paramagnetic depairing, this continuum approximation is safely valid if $H_P(0) \ll B_{\rm cr}$, or 
\begin{equation}
\gamma \ll \frac{1}{2e H_P d^2} = 
\frac{1.3 \xi_0^2}{d^2} \alpha_M^{({\rm 2D})},
\label{paulidame}
\end{equation}
where $\alpha_M^{({\rm 2D})} = \sqrt{2} 
H_{\rm orb}^{({\rm 2D})}(0)/H_P(0)$, and 
$H_{\rm orb}^{({\rm 2D})}(0)
% = 0.56/(2 e \xi_0^2)
$ is the orbital limiting field in 2D limit for fields perpendicular to the layers. The above inequality means that an increase of $\gamma$ {\it competes} with an enhanced 
paramagnetic depairing. In fact, eq.(\ref{paulidame}) implies that, as the paramegnetic depairing is stronger under a fixed anisotropy, the FFLO state just below $H_{c2}(T)$ tends to enter not the nearly 2D region in $B > B_{\rm cr}$ but the anisotropic 3D regime below $B_{\rm cr}$ in which repeated structural transitions between the Josephson vortex lattices occur. Nevertheless, the transitions between the ordinary Josephson vortex lattices are not visible in most of quasi 2D materials, and the layer structure may be treated as an anisotropic 3D-like medium for most purposes as far as the intrinsic pinning effect of vortices does not become essential \cite{comUji}. 
Then, the difference $\varepsilon_{{\bf p}_+} - \varepsilon_{{\bf p}_-}$ of the quasiparticle energy may be replaced by ${\bf w}\cdot{\bf k}$ even in layered systems. Consequently, the quadratic term of the GL free energy density is given by 
%%%%%
\begin{equation}
{\cal F}_2 
= 	\frac{1}{V} \int d^3 r {\Delta}^*({\bf r})
	\left( \frac{1}{|g|}-\hat{K}_{2}({\bf \Pi}) \right) 
	{\Delta}({\bf r}), 
\label{F21}
\end{equation}
%%%%%%
where ${\bf w}$ is the velocity field on the FS, 
${\bf \Pi}=-{\rm i} \nabla + 2 e {\bf A}$, and an appropriate gauge transformation has been performed to make the gauge field parallel to 
the $z$-direction. 
The operator $\hat{K}_2$ is given by 
%%%%%
\begin{eqnarray}
 \hat{K}_2 ({\bf \Pi}) &=& \frac{T}{2} \sum_{\varepsilon, \sigma} \int_{\bf p} |{\hat \Delta}_{\bf p}|^2 G_{\varepsilon, \sigma}({\bf p}) G_{-\varepsilon, -\sigma}(-{\bf p}+{\bf \Pi}) \nonumber \\
&=&
 \pi N(0) T \sum_{\varepsilon, \sigma} \biggl\langle |{\hat \Delta}_{\bf p}|^2 \frac{{\rm i} {\rm sgn}(\varepsilon_n)}{2({\rm i}\varepsilon_n + \sigma \mu H) - {\bf w}\cdot{\bf \Pi}} \biggr\rangle_{\rm FS}  \nonumber \\ 
 &=& N(0) \int_0^\infty d\rho \, f(\rho) \biggl\langle |{\hat \Delta}_{\bf p}|^2 \exp({\rm i} \, T_c^{-1} \rho {\bf w}\cdot{\bf \Pi}) \biggr\rangle_{\rm FS},
\nonumber \\ \label{K21}
\end{eqnarray}
%%%%%
Here, $\langle \,\,\, \rangle_{\rm FS}$ denotes the average over the Fermi surface (FS), $t=T/T_c$, $N(0) = N_2(0)/d$ with the density of states $N_2(0)$ per spin in 2D case, 
\begin{equation}
f(\rho) = \frac{2 \pi \, t}{{\rm sinh}(2 \pi t \rho)} \, {\rm cos}\biggl(\frac{2 \mu B \, \rho}{T_c} \biggr), 
\label{f}
\end{equation}
and the identity 
\begin{equation}
\frac{1}{D} = \int_0^\infty d\rho \exp(-\rho D)
\label{parameterint}
\end{equation}
was used in obtaining the last equality of eq.(\ref{K21}). 
Similarly, the 4-th order (quartic) term and the 6-th order one of the GL free energy density are written as 
\begin{eqnarray}
{\cal F}_4 &=& \frac{1}{2 V} \int d^3r \, {\hat K}_4({\bf \Pi}_j) \, \Delta^*({\bf r}_1) \, \Delta^*({\bf r}_3) \, \Delta({\bf r}_2) \, \Delta({\bf r}_4)|_{{\bf r}_j \to {\bf r}}, \nonumber \\ 
{\cal F}_6 &=& \frac{1}{3 V} \int d^3r \, {\hat K}_6({\bf \Pi}_j) \, \Delta^*({\bf r}_1) \, \Delta^*({\bf r}_3) \, \Delta^*({\bf r}_5) \, \Delta({\bf r}_2) \, \, \Delta({\bf r}_4) \, \Delta({\bf r}_6)|_{{\bf r}_j \to {\bf r}}, 
\label{F4}
\end{eqnarray}
where 
\begin{eqnarray}
{\hat K}_4 &=& \frac{T}{2} \sum_{\varepsilon,\sigma} \int_{\bf p} |{\hat \Delta}_{\bf p}|^4 \, G_{\varepsilon, \sigma}({\bf p}) \, G_{-\varepsilon, -\sigma}(-{\bf p}+{\bf \Pi}_1^*) G_{-\varepsilon, -\sigma}(-{\bf p}+{\bf \Pi}_2) G_{\varepsilon, \sigma}({\bf p}+{\bf \Pi}_3^* - {\bf \Pi}_2) \nonumber \\ 
&=& 2 \pi N(0) T \sum_{\varepsilon, \sigma} \biggl\langle \frac{-{\rm i} {\rm sgn}(\varepsilon) \, |{\hat \Delta}_{\bf p}|^4}{d_1 d_2 d_3} 
\biggr\rangle_{\rm FS}, 
\label{K41}
\end{eqnarray}
and 
\begin{eqnarray}
{\hat K}_6 &=& - \frac{T}{2} \sum_{\varepsilon,\sigma} \int_{\bf p} |{\hat \Delta}_{\bf p}|^6 \, G_{\varepsilon, \sigma}({\bf p}) \, G_{-\varepsilon, -\sigma}(-{\bf p}+{\bf \Pi}_1^*) G_{-\varepsilon, -\sigma}(-{\bf p}+{\bf \Pi}_6) \nonumber \\
&\times& G_{\varepsilon, \sigma}({\bf p}-{\bf \Pi}^*_1 + {\bf \Pi}_2) \, G_{-\varepsilon, -\sigma}(-{\bf p}+{\bf \Pi}^*_1 + {\bf \Pi}^*_3 - {\bf \Pi}_2) \, G_{\varepsilon, \sigma}({\bf p}-{\bf \Pi}_6 + {\bf \Pi}^*_5) \nonumber \\
&=& \pi N(0) T \sum_{\varepsilon,\sigma} \biggl\langle \frac{(-{\rm i} {\rm sgn}(\varepsilon)) |{\hat \Delta}_{\bf p}|^6} {d_1 d_2 d_3 d_4 d_5 d_6}  \frac{\sum_{i, {\rm perm}} d_i d_{i+1} d_{i+2} d_{i+3}}{(d_1+d_3-d_2)(d_3+d_5-d_4)(d_5+d_1-d_6)} \biggr\rangle_{\rm FS}. 
\label{K61}
\end{eqnarray}
Here, $\sum_{\rm perm} d_i d_{i+1} d_{i+2} d_{i+3} = d_1 d_2 d_3 d_4 + \cdot \cdot \cdot + d_6 d_1 d_2 d_3$, $d_j=2({\rm i} \varepsilon_n + \sigma \mu B) - {\bf w}\cdot{\bf \Pi}_j$ for an even $j$, and $d_j=2({\rm i} \varepsilon_n + \sigma \mu B) - {\bf w}\cdot{\bf \Pi}_j^*$ for an odd $j$. Although the above expression of the 6-th order term is apparently different from the corresponding one in \cite{AI}, it can be numerically checked that both of them are the same as each other. By using the identity (\ref{parameterint}) again, we obtain 
\begin{eqnarray}
{\hat K}_4 &=& \frac{2}{T_c^2} \, N(0) \int\Pi_{j=1}^3 d\rho_j \, f\bigl(\sum_{j=1}^3 \rho_j \bigr) \biggl\langle |{\hat \Delta}_{\bf p}|^4 \, \exp\biggl[\frac{\rm i}{T_c}( \, \rho_1 {\bf w}\cdot{\bf \Pi}_1^* + \rho_2 {\bf w}\cdot{\bf \Pi}_2 + \rho_3 {\bf w}\cdot{\bf \Pi}_3^* \, ) \biggr] \biggr\rangle_{\rm FS}, \nonumber \\
{\hat K}_6 &=& - \frac{6}{T_c^4} \, N(0) \int\Pi_{j=1}^5 d\rho_j \, f\bigl(\sum_{j=1}^5 \rho_j \bigr) \biggl\langle |{\hat \Delta}_{\bf p}|^6 \, \exp\biggl[\frac{\rm i}{T_c}( \, \rho_1 {\bf w}\cdot{\bf \Pi}_5^* + \rho_2 {\bf w}\cdot{\bf \Pi}_6 \nonumber \\
&+& \rho_3 {\bf w}\cdot({\bf \Pi}_3^*+{\bf \Pi}_1^* - {\bf \Pi}_2) + \rho_4 {\bf w}\cdot({\bf \Pi}_3^*+{\bf \Pi}_5^*-{\bf \Pi}_4) + \rho_5 {\bf w}\cdot({\bf \Pi}_5^*+{\bf \Pi}_1^*-{\bf \Pi}_6) \, ) \biggr] \biggr\rangle_{\rm FS}
\label{K46}
\end{eqnarray}

Hereafter, the order parameter field $\Delta$ will be decomposed into the LLs. When the FFLO state modulating along ${\bf H}$ is formed, no additional spatial variation is induced in the $y$-$z$ plane perpendicular to ${\bf H} \parallel {\hat x}$ except that due to the vortex structure, and hence, the $y$ and $z$ dependences of $\Delta$ of the FFLO state in equilibrium is, as in the Abrikosov state, well described in the lowest ($n=0$) LL. Nevertheless, in ${\bf H} \parallel {\hat x}$ parallel to the layers, the anisotropy $\gamma$ in the $y$-$z$ plane between the coherence lengths needs to be determined to consistently define creation and annihilation operators $\Pi_\pm$ for the LLs. 
To determine $\gamma$, we follow its derivation in the {\it conventional} GL region and focus on the quadratic term in ${\bf \Pi}$ and ${\bf \Pi}^*$ which is proportional to 
\begin{eqnarray}
\langle (w_y \Pi_y &+& w_z \Pi_z)^2 \rangle_{\rm FS} = \gamma^{-1} \langle w_y^2 \rangle_{\rm FS} \, [ \, r_B^{-2} + (\gamma^{1/2} \Pi_y - {\rm i} \gamma^{-1/2} \Pi_z)(\gamma^{1/2} \Pi_y + {\rm i} \gamma^{-1/2} \Pi_z) \, ], 
\label{anisotropy}
\end{eqnarray}
where $\gamma=\sqrt{\langle w_y^2 \rangle/\langle w_z^2 \rangle}$. Here and below, we have chosen the gauge ${\bf A} = -B y {\hat z}$ leading to $\Pi_- \Pi_+ - \Pi_+ \Pi_- = 1$ where $\Pi_\pm = r_B(\gamma^{-1/2} \Pi_z \pm {\rm i} \gamma^{1/2} \Pi_y)/\sqrt{2}$ are the creation and annihilation operators of the LLs satisfying $\Pi_- \varphi_0(y,z)=0$, and $\varphi_n(y,z)$ is a basis function in the $n$-th LL. According to ${\cal H}_0$ defined above, the velocity field on the FS is given  by 
\begin{eqnarray}
{\bf w}(\phi, k_z) &=& v_F ( 1 - {\tilde J}(1 - {\rm cos}(p_z d)))^{1/2} \, ({\rm cos}(\phi) {\hat x} + {\rm sin}(\phi) {\hat y}) + J \, d \, {\rm sin}(p_z d) {\hat z}
\label{Fermivel}
\end{eqnarray}
where ${\tilde J}=2 m J/p_F^2$. In this case, we have $\gamma=2 \sqrt{1 - {\tilde J}}/(\pi {\tilde J})$. For this ${\bf w}$, the averaging over the FS is defined by 
\begin{equation}
\langle M \rangle_{\rm FS} = \int_0^{2 \pi} \frac{d\phi}{2 \pi} \int_{-\pi}^{\pi} \frac{d(p_z d)}{2 \pi} \, \, M.
\label{FSaver}
\end{equation}

Next, the operation ${\tilde \varphi_n}(y,z) \equiv \exp({\rm i} \, \rho T_c^{-1} {\bf w}\cdot{\bf \Pi}) \varphi_n(y,z)$, necessary to make the expressions of ${\cal F}_{2n}$ tractable for numerical calculations, will be examined. This is most easily accomplished in terms of the corresponding coherent state \cite{PAL} $\exp(-|\sigma|^2/2)\sum_{n \geq 0} \sigma^n \varphi_n/\sqrt{n !}$. Examining the action of $\exp({\rm i} \, T_c^{-1} {\bf w}\cdot{\bf \Pi})$ to this coherent state, we obtain  
\begin{eqnarray}
{\tilde \varphi_n}(y,z) &\equiv& \exp\biggl({\rm i} \, \frac{\rho}{T_c} {\bf w}\cdot{\bf \Pi} \biggr) \varphi_n(y,z) \nonumber \\
&=& \frac{{\cal N}_0}{\sqrt{2!}} \, \exp\biggl( \frac{\rho^2}{2}(\mu^2 - |\mu|^2) \biggr) \biggl(\rho(\mu - \mu^*) + \frac{\partial}{\partial (\rho \mu)} \biggr)^n \, \exp\biggl(-{\rm i}p{\hat z} - \frac{({\hat y}+\sqrt{2}\rho \mu + p)^2}{2} \biggr) 
\label{opeLL}
\end{eqnarray}
for $n \leq 2$, where ${\hat y}= \gamma^{-1/2} y$, ${\hat z}=\gamma^{1/2} z$, 
\begin{equation}
\mu=\frac{\gamma^{-1/2} w_y 
+ {\rm i} \gamma^{1/2} w_z}{\sqrt{2} \, \, r_B T_c},
\label{mu}
\end{equation}
and the corresponding $n=0$ LL state is 
\begin{equation}
\varphi_0(y,z) = {\cal N}_0 \exp\biggl(-{\rm i}p{\hat z} - \frac{({\hat y}+ p)^2}{2} \biggr). 
\end{equation}

At this stage, it is straightforward to study properties in equilibrium of an FFLO vortex state. First, as already mentioned, we assume the FFLO state in equilibrium to be described in the lowest ($n=0$) LL where no nodal points or lines except the field-induced vortices of $|\Delta|$ are present in the $y$-$z$ plane perpendicular to ${\bf H}$. Instead, nodal planes perpendicular to ${\bf H}$ are periodically formed. If the $Q$-dependence of the free energy is incorporated only from the quadratic term ${\cal F}_2$, the transition line between the LO state and the Abrikosov state is the same as that between the FF and Abrikosov states, where $Q$ is the wavenumber corresponding to the period between the FFLO nodal planes. However, once the $Q$ dependence of the free energy from the higher order terms, ${\cal F}_k$ ($k \geq 4$), is considered, as pointed out elsewhere \cite{IA} in the ${\bf H} \parallel {\hat z}$ case, the transition temperature between the LO and Abrikosov states is higher than that between the FF and Abrikosov states. For this reason, we will not consider a possibility of appearance of the FF state. Hereafter, the equilibrium order parameter $\Delta_e$ in the FFLO state is assumed to take the form
\begin{equation}
\Delta_e({\bf r}) = \sqrt{2} \, T_c \, \alpha_e \, \Psi_{\rm A}(y,z) \, {\rm cos}(Qx)
\label{eqFFLO}
\end{equation}
with the normalization condition $\langle |\Delta_e|^2 \rangle_{\rm sp} = T_c^2$, where $\langle \,\,\,\, \rangle_{\rm sp}$ implies the spatial average, and $\Psi_{\rm A}$ is the Abrikosov solution of the vortex lattice in $n=0$ LL. Then, in equilibrium,  the mean field free energy density of the FFLO state takes the form 
\begin{eqnarray}
\frac{{\cal F}_e}{N(0) T_c^2} &=& a_0(q) \alpha_e^2 + \frac{V_4(q)}{2} \alpha_e^4 + \frac{V_6}{3} \alpha_e^6 \nonumber \\
&=& c^{(0)}(\alpha_e) + c^{(2)}(\alpha_e) q^2 + c^{(4)}(\alpha_e) q^4.
\label{fe}
\end{eqnarray}
Hereafter, the dimensionless wavenumber of the FFLO modulation will be defined as 
\begin{equation}
q = Q \, r_B \gamma^{1/2}. 
\label{scaledQ}
\end{equation} 
The coefficients in eq.(\ref{fe}) are given by 
\begin{eqnarray}
a_0(q) &=& a_0(0) + a_0^{(2)} q^2 - a_0^{(4)} q^4, \nonumber \\
V_4(q) &=& V_4(0) - V_4^{(2)} q^2 + V_4^{(4)} q^4. 
\label{GLcoef1}
\end{eqnarray}
Hereafter, the $q$ dependence of $V_6$ will be neglected. This simplification is in part based on the fact that it has been verified \cite{IA} that the $q$ dependences of $V_6$ are unimportant even quantitatively for the position of the FFLO to Abrikosov transition in ${\bf H} \parallel {\hat z}$. Further, to study systematically possible phase diagrams including FFLO states, inclusion of $q$-dependences in higher order terms requires a difficult numerical task. On the other hand, if even the $q$ dependence of $V_4$ is neglected, the $B$-$T$ region in which the FFLO state can appear is highly overestimated, and, as is seen in sec.III, a fictitious tilt instability of the FFLO state occurs. Therefore, for the practical purposes, neglecting $q$ dependences of $V_6$ and keeping the corresponding ones of $V_4(q)$ is a convincing approach. Of course, when using eq.(\ref{fe}), it is necessary to verify the conditions $V_6 > 0$ and $c^{(4)} > 0$ which justify the use of ${\cal F}_e$ truncated at the O($|\Delta|^6$) and O($q^4$) terms. 

The onset temperature $T_0$ at which the mean field $H_{c2}$-transition becomes discontinuous is given as the position at which $V_4(q)$ becomes negative upon cooling while $V_6 > 0$, and a second order transition line $H_{\rm FFLO}(T)$ is determined as the line on which $c^{(2)}(\alpha_{\rm e})$ in $B < H_{c2}$ becomes negative on cooling, while $c^{(4)}(\alpha_{\rm e}) > 0$, .
Then, by minimizing ${\cal F}_e$ with respect both to $q$ and $\alpha_e$, 
$\alpha_e^2$ is determined by  
\begin{eqnarray}
\alpha_e^2 (q) &=& \frac{-V_4(q) + \sqrt{(V_4(q))^2 - 4 a_0(q) V_6}}{2 V_6}, 
\label{equilalpha}
\end{eqnarray}
while $q=0$ if $a_0^{(2)} - V_4^{(2)} (\alpha_e(q))^2/2 > 0$, and 
\begin{eqnarray}
q^2 = \frac{-a_0^{(2)} + V_4^{(2)} (\alpha_e(q))^2/2}{2(-a_0^{(4)}+(\alpha_e(q))^2 V_4^{(4)}/2)},
\label{equilq} 
\end{eqnarray}
otherwise. Below $T_0$, the discontinuous $H_{c2}$-transition (i.e., first order mean field SC transition) occurs when 
\begin{equation}
a_0(q)=\frac{3}{16} \, \frac{(V_4(q))^2}{V_6} 
\label{fot}
\end{equation}
irrespective of the minimized value of $q$. 

By applying eq.(\ref{eqFFLO}) to ${\cal F}_j$, it is straightforward to derive the coefficients in eq.(\ref{fe}) if eq.(\ref{parameterint}) is 
repeatedly used. Consequently, they are given by 
\begin{eqnarray}
a_0(0) &=& \frac{1}{2}{\rm ln}(h) + \int_0^\infty d\rho \biggl[ \frac{1}{\rho} \exp\biggl(-\frac{\pi^2 \xi_0^2 \rho^2}{r_B^2} \biggr) - f(\rho) \biggl\langle |{\hat \Delta}_p|^2 \exp\biggl(-\frac{|\mu|^2 \rho^2}{2} \biggr) \biggr\rangle_{\rm FS} \biggr], 
\nonumber \\
a_0^{(2)} &=& \int_0^\infty d\rho \, f(\rho) \rho^2 \biggl\langle \, ({\rm Re} \mu)^2 {\rm cot}^2\phi \, |{\hat \Delta}_p|^2 \exp\biggl(-\frac{|\mu|^2 \rho^2}{2} \biggr) \biggr\rangle_{\rm FS}, \nonumber \\
a_0^{(4)} &=& \frac{1}{6} \int_0^\infty d\rho  \, f(\rho) \rho^4 \biggl\langle \, ({\rm Re} \mu)^4 \, {\rm cot}^4\phi \, |{\hat \Delta}_p|^2 \exp\biggl(-\frac{|\mu|^2 \rho^2}{2} \biggr) \biggr\rangle_{\rm FS}, \nonumber \\
V_4(0) &=& 3 \int_0^\infty \Pi_{j=1}^3 d\rho_j \, f\biggl(\sum_{j=1}^3 \rho_j \biggr) \biggl\langle |{\hat \Delta}_p|^4 \exp\biggl(-\frac{1}{2} \biggl(-\frac{1}{2} R_{24} + R_{14} \biggr) \biggr) \, {\rm cos}(I_4) \biggr\rangle_{\rm FS}, \nonumber \\
V_4^{(2)} &=& 3 \int_0^\infty \Pi_{j=1}^3 d\rho_j \, f\biggl(\sum_{j=1}^3 \rho_j \biggr) \biggl(\sum_{j=1}^3 \rho_j^2 - \frac{1}{3} \sum_{i \neq j} (-1)^{i+j} \, \rho_i \rho_j \biggr) \nonumber \\
&\times& \biggl\langle \, ({\rm Re}\mu)^2 {\rm cot}^2\phi \, |{\hat \Delta}_p|^4 \exp\biggl(-\frac{1}{2} \biggl(-\frac{1}{2} R_{24} + R_{14} \biggr) \biggr) \, {\rm cos}(I_4) \biggr\rangle_{\rm FS}, \nonumber \\
V_4^{(4)} &=& \frac{1}{2} \int_0^\infty \Pi_{j=1}^3 d\rho_j \, f\biggl(\sum_{j=1}^3 \rho_j \biggr) \biggl[ \, \sum_{j=1}^3 \rho_j^4 + \sum_{i \neq j} ( 3 \rho_i^2 \rho_j^2 - 2 (-1)^{i+j} \rho_i \rho_j (\rho_{6-i-j})^2 - \frac{4}{3}(-1)^{i+j} \rho_i \rho_j^3) \biggr] \nonumber \\
&\times& \biggl\langle \, ({\rm Re}\mu)^4 {\rm cot}^4\phi \, |{\hat \Delta}_p|^4  \exp\biggl(-\frac{1}{2} \biggl(-\frac{1}{2}R_{24} + R_{14} \biggr) \biggr) \, {\rm cos}(I_4) \biggr\rangle_{\rm FS}, \nonumber \\
V_6 &=& -15  \int \Pi_{j=1}^5 d\rho_j \, f\biggl(\sum_{k=1}^5 \rho_k \biggr) \biggl\langle |{\hat \Delta_p}|^6 \exp\biggl( -\frac{1}{2}(R_{16}+R_{26}) \biggr) \, {\rm cos}(I_6) \biggr\rangle_{\rm FS},
\label{GLcoef2}
\end{eqnarray}
where $h=B/H_{\rm orb}^{({\rm 2D})}(0)
% = 1.79 \xi_0^2/r_B^2
$, and 
\begin{eqnarray}
R_{14} &=& |\mu|^2 (\sum_{j=1}^3 \rho_j^2 + \rho_2(\rho_3+\rho_1)), 
\nonumber \\
R_{24} &=& {\rm Re}(\mu^2) (\rho_2^2 + (\rho_3-\rho_1)^2), \nonumber \\
I_4 &=& \frac{{\rm Im}(\mu^2)}{4} (\rho_2^2 - (\rho_3-\rho_1)^2), \nonumber \\
R_{16} &=& |\mu|^2 \biggl(e_1+e_2+e_3+\frac{2}{3}e_4 e_5 \biggr), \nonumber \\
R_{26} &=& {\rm Re}\mu^2 \biggl(e_1+e_2+e_3-\frac{e_4^2+e_5^2}{3}-\frac{2}{3} (e_6+e_7+e_8+e_9) \biggr), \nonumber \\ 
I_6 &=& \frac{{\rm Im}(\mu^2)}{4} (e_1+e_2-e_3+\frac{e_5^2-e_4^2}{3} + \frac{2}{3}(e_8+e_9-e_6-e_7)) \nonumber \\ 
e_1 &=& (\rho_3+\rho_5)^2 + (\rho_3+\rho_4)^2, \nonumber \\
e_2 &=& (\rho_1+\rho_4+\rho_5)^2, \nonumber \\
e_3 &=& \rho_3^2 + \rho_4^2 + (\rho_2-\rho_5)^2, \nonumber \\
e_4 &=& \rho_1+2(\rho_3+\rho_4+\rho_5), \nonumber \\
e_5 &=& \rho_2-\rho_3-\rho_4-\rho_5, \nonumber \\
e_6 &=& (\rho_4-\rho_5)^2 + (\rho_1+\rho_5-\rho_3)^2, \nonumber \\
e_7 &=& (\rho_1+\rho_4-\rho_3)^2, \nonumber \\
e_8 &=& (\rho_3-\rho_4)^2 + (\rho_2+\rho_3-\rho_5)^2, \nonumber \\
e_9 &=& (\rho_2+\rho_4-\rho_5)^2. 
\end{eqnarray}
In obtaining $a_0(0)$, the interaction strength $|g|$ has been eliminated under such a condition that, in ${\bf H}$ perpendicular to the layers, the operation $|g|^{-1} - {\hat K}_2$ at $T=0$ and in the absence of the paramagnetic effect vanishes at $H_{\rm orb}^{({\rm 2D})}(0)$.  

It should be noted that, for any FS with anisotropy in the $y$-$z$ plane, the $n=2$ LL mode couples to the $n=0$ LL mode of $\Delta$ in high fields, and hence, that the expressions of GL coefficients given above are, strictly speaking, insufficient. This coupling inevitably occurs except in the conventional GL region valid in lower fields, where the gradient terms are kept only up to O(${\bf \Pi}^2$), and the $y$ and $z$ dependences of expressions were isotropized in determining $\gamma$ (see the description around eq.(\ref{anisotropy})). Expressions of coefficients in the GL quadratic term related to the $n=2$ LL modes are given in Appendix A. For the anisotropy values ($\gamma \leq 5.5$) used in our numerical calculations, however, this coupling was quantitatively negligible, and thus, the coefficients in eq.(\ref{GLcoef2}) 
are used in obtaining numerical results in sec.III. 

Now, let us turn to explaining how to describe tilt deformations in the FFLO state with nodal planes perpendicular to ${\bf H}$. In the present parallel field configuration, the in-plane vortex tilt unaffected by the intrinsic pinning effect of the layering is expressed as a $x$-dependent vortex displacement ${\bf u} = u{\hat y}$ related to a vortex flow in the $y$-direction. In a vortex lattice in equilibrium described in the $n=0$ LL, such a tilt deformation accompanied by a vortex flow is, consistently with the vanishing of the corresponding static superfluid rigidity, expressed as a fluctuation in the next lowest ($n=1$) LL of $\Delta$. \cite{IR92} Such a relation between the $n=1$ LL mode and the vortex displacement will be reviewed in Appendix B within the conventional GL region valid in low fields and near $T_c$. Through the analysis shown there and in Ref.\cite{IR92}, it is convincingly understood that the following points are valid beyond the conventional GL region. First, in examining the elastic deformations of the Abrikosov lattice, the energy gap between the $n=1$ and $n=0$ LLs is lost due to the magnetic screening, i.e., a gauge field fluctuation coupling to the vortex motion, and this disappearance of the mass gap is equivalent to the vanishing of the static superfluid rigidity, $\Upsilon_{s \, \perp}=0$, for a phase twist perpendicular to ${\bf H}$. The resulting main term of the vortex tilt modulus is the magnetic energy $B^2/(4 \pi)$ which is insensitive to the details of the SC state. Clearly, this result that the main term of $C_{44}$ becomes insensitive to the details of the SC state as a consequence of the vanishing of $\Upsilon_{s \, \perp}$ holds true in the FFLO state modulating along ${\bf H}$. On the other hand, according to the results in the conventional GL case in Appendix B, the remaining term $\delta C_{44} = C_{44} - B^2/(4 \pi)$ arises directly from the gradient ($\partial_x$) term of the resulting GL action regardless of the magnetic screening. That is, as far as focusing on $\delta C_{44}$, we can work in type II limit with no fluctuation of the gauge field. Then, consistently with eq.(\ref{eqFFLO}), the SC order parameter field with tilt deformations of the vortices and of the nodal planes included should take the form 
\begin{equation}
\Delta = \Delta_e(y,z) + \delta \Delta({\bf r}) = \sqrt{2} \, \alpha_e \, T_c \, (\varphi_0(y,z) + \delta a_1(x) \varphi_1(y,z)) \, 
{\rm cos}(Qx+T_c s(x,y)/v_F), 
\label{alphae}
\end{equation}
where $s(x,y)$ is the displacement of the nodal planes, and the amplitude $\delta a_1(x)$ of the $n=1$ LL fluctuation is identified with the vortex displacement ${\bf u} = u{\hat y}$ parallel to the layers in the manner 
\begin{equation}
\delta a_1(x) = \frac{u(x)}{\sqrt{2 \gamma} \, r_B},
\label{n=1fluc}
\end{equation}
as explained in Appendix B. 
If $s(x,y)=0$, $\delta C_{44}$ is obtained as the coefficient of $(\partial_x u)^2$ term. 

Hereafter, the elastic constants $E_j$ ($j=1$, $2$, and $3$) introduced in sec.1 will be expressed in the manner 
\begin{eqnarray}
E_1 &=& 2 \pi^2 \frac{\xi_0^2}{\gamma \, r_B^2} \alpha_e^2 \biggl( L_1 + \frac{\alpha_e^2}{2} {\overline L}_1  \biggr), \nonumber \\
E_2 &=& \alpha_e^2 \biggl( L_2 + \frac{\alpha_e^2}{2} {\overline L}_2 \biggr), \nonumber \\
E_{3} &=& \sqrt{2} \pi \frac{\xi_0}{\gamma^{1/2} \, r_B} \alpha_e^2 \biggl( L_3 + \frac{\alpha_e^2}{2} {\overline L}_3 
\biggr).
\label{elascon}
\end{eqnarray}
To first obtain the contributions $L_j$ ($j=1$, $2$, and $3$) from the quadratic GL term ${\cal F}_2$, let us consider the following quantity 
\begin{eqnarray}
\langle 0|{\tilde n} \rangle &\equiv& 2 \biggl\langle \biggl\langle \varphi^*_0(y,z) \, {\rm cos}(Qx+ T_c s(x,y)/v_F) \, \exp\biggl({\rm i}\frac{\rho}{T_c}({\bf w}\cdot{\bf \Pi}) \biggr) \nonumber \\
&\times& \delta a_n(x) \varphi_n(y,z) \, {\rm cos}(Qx+ T_c s(x,y)/v_F) \biggr\rangle_{\rm FS} \biggr\rangle_{\rm sp}
\label{matrixelem1}
\end{eqnarray}
($n=0$ or $1$) appearing commonly in ${\cal F}_j$ ($j=2$, $4$, and $6$), where $\delta a_0 \equiv 1$ is assumed. First, using the identity $\exp(A+B)=\exp(A) \exp(B) \exp(-(AB-BA)/2)$, valid when $AB-BA$ is a constant, the expression 
\begin{eqnarray}
\exp\biggl(\frac{\rm i}{T_c} \rho  {\bf w}\cdot{\bf \Pi} \biggr)  \delta a_n(x) \varphi_n(y,z) \, {\rm cos}(Qx+ T_c s(x,y)/v_F) 
\label{matrixex1}
\end{eqnarray}
will be written as 
\begin{eqnarray}
\delta a_n(x+T_c^{-1} w_x \rho) \, {\rm cos}( \, Q(x+T_c^{-1}\rho w_x) 
+ T_c \, \, s(x+T_c^{-1}\rho w_x, y+T_c^{-1}\rho w_y) \, ) \, \exp\biggl(\frac{\rm i}{T_c} \rho (w_y \Pi_y+w_z \Pi_z) \biggr) \varphi_n(y,z). 
\label{matrixex2}
\end{eqnarray}
Then, the average on $x$ over the scale $2 \pi/Q$ will be performed prior to all of the spatial averages by assuming a slow variation of $s(x)$ in $x$, and thus, $2 {\rm cos}^2(Qx+s(x,y))$ may be replaced by unity. Further, using eqs.(\ref{matrixex1}) and (\ref{matrixex2}) and keeping only terms remaining finite after the momentum average on the FS, we find 
\begin{equation}
\langle 0|{\tilde 0} \rangle = \langle {\rm cos}(\rho Q w_x/T_c) \, \langle {\rm cos}(\rho w_y \partial_y s/v_F) \, {\rm cos}(\rho w_x \partial_x s/v_F) \, \exp(-\rho^2 |\mu|^2/2) \rangle_{\rm sp} \rangle_{\rm FS}, 
\label{matrix00}
\end{equation}
and 
\begin{eqnarray}
\langle 0|{\tilde 1} \rangle &=& \biggl\langle \biggl\langle \rho \frac{w_x}{T_c} \partial_x \delta a_1(x) \biggl(-\frac{w_y}{v_F} \rho \partial_y s(x,y) \biggr) \, {\rm sin}\biggl(\rho Q \frac{w_x}{T_c} \biggr) (- \rho \mu^*) \exp\biggl(-\frac{\rho^2 |\mu|^2}{2} \biggr) \biggr\rangle_{\rm sp} \biggr\rangle_{\rm FS} \nonumber \\
&=& \frac{2 \pi^2 \rho^3}{\gamma} \, \biggl\langle \biggl(\frac{\xi_0 \, w_y}{r_B \, v_F} \biggr)^2 \frac{w_x}{v_F} {\rm sin}\biggl(\rho \frac{w_x}{T_c} Q \biggr) \, \exp\biggl(-\frac{\rho^2 |\mu|^2}{2} \biggr) \biggr\rangle_{\rm FS}  \langle \partial_x u \, \partial_y s \rangle_{\rm sp}.
\label{matrix01}
\end{eqnarray}
The latter is valid up to the harmonic order in $s$ and $u$. Then, $L_j$'s are easily obtained in terms of the above expressions and are given by 
\begin{eqnarray}
L_{1} &=& \! \int_0^\infty \! d\rho_0 \biggl\langle |{\hat \Delta}_{\hat p}|^2 f(\rho_0) \biggl(\frac{w_x \rho_0}{v_F} \biggr)^2 (1 - \rho_0^2 |\mu|^2)  \exp\biggl(-\frac{\rho_0^2|\mu|^2}{2} \biggr) \, {\rm cos}(0) \biggr\rangle_{\rm FS}, \\ \nonumber
L_{2} &=& \! \int_0^\infty \! d\rho_0 \biggl\langle |{\hat \Delta}_{\hat p}|^2 f(\rho_0) \biggl(\frac{w_y \rho_0}{v_F} \biggr)^2 \!\! \exp\biggl(-\frac{\rho_0^2|\mu|^2}{2} \biggr) \, {\rm cos}(0) \biggr\rangle_{\rm FS}, \\ \nonumber
L_{3} &=& \int_0 d\rho_0 \biggl\langle |{\hat \Delta}_{\hat p}|^2 f(\rho_0) \, {\rm Re}(\mu) \, \rho_0 \frac{w_y \rho_0}{v_F} \frac{w_x \rho_0}{v_F} \exp\biggl(-\frac{\rho_0^2|\mu|^2}{2} \biggr) \, {\rm sin}(0) \biggr\rangle_{\rm FS}, 
\label{Lj}
\end{eqnarray} 
where 
\begin{eqnarray}
{\rm cos}(n) &=& {\rm cos}\biggl(\sqrt{2} \, q \, \rho_n \, {\rm Re}(\mu) {\rm cot}\phi \, \biggr), \nonumber \\
{\rm sin}(n) &=& {\rm sin}\biggl(\sqrt{2} \, q \, \rho_n \, {\rm Re}(\mu) {\rm cot}\phi \, \biggr). 
\label{cnsn}
\end{eqnarray}

It might be natural to discuss the elastic deformation of the vortex lattice based on these $L_j$'s without including the contributions from ${\cal F}_4$. However, $L_2$ itself is found not to lead to a qualitatively reasonable result of $E_2$ in the FFLO state: As is shown later in Fig.4, the resulting tilt rigidity $L_2$ {\it of the nodal planes} often becomes negative. This result, suggestive of an instability of the FFLO state modulating along ${\bf H}$, is an artifact due to the neglect of contributions to $E_2$ from the higher order terms, ${\cal F}_{4}$, of the GL free energy. Hereafter, consistently with the neglect of $q$-dependences in $V_6$ in eq.(\ref{fe}), $E_n$ will be expressed, as in eq.(\ref{elascon}), as the sum of the contributions of ${\cal F}_2$ and ${\cal F}_4$ terms of the GL free energy. Derivation of the contributions ${\overline L}_j$ to the elastic moduli from ${\cal F}_4$ is lengthy but straightforward once using the expressions (\ref{matrixex1}) and (\ref{matrixex2}), and they are expressed by 
\begin{eqnarray}
{\overline L}_1 &=& \int_0 \Pi_{j=1}^3 d\rho_j (\rho_1+\rho_2)^2 \, f\biggl(\sum_{j=1}^3 \rho_j \biggr) \biggl\langle |{\hat \Delta}_p|^4 \, \biggl(\frac{w_x}{v_F} \biggr)^2 \biggl(3 \Pi_{j=1}^3 {\rm cos}(j) + \frac{1}{2} \sum_{i \neq j} (-1)^{i+j} {\rm sin}(i) {\rm sin}(j) {\rm cos}(6-i-j) \biggr) \nonumber \\
&\times& \biggl[\biggl( -1 + \frac{1}{4}[ \, (3\rho_1+\rho_2-\rho_3)(3\rho_2+\rho_1+\rho_3)(|\mu|^2 + {\rm Re}(\mu^2)) + [(3\rho_1-\rho_2-\rho_3)(3\rho_2-\rho_3-\rho_1) + 4(\rho_1-\rho_2)^2 \nonumber \\ 
&+& 4\rho_3(\rho_1+\rho_2) ](|\mu|^2-{\rm Re}(\mu)^2) \, ] \biggr) \, {\rm cos}(I_4) + \frac{1}{2} {\rm Im}(\mu)^2((\rho_2+\rho_3)^2-\rho_1^2) \, {\rm sin}(I_4) \biggr] \exp\biggl(-\frac{1}{2}\biggl(-\frac{1}{2}R_{24}+R_{14} \biggr) \biggr) \biggr\rangle_{\rm FS}, \nonumber \\
{\overline L}_2 &=& \int_0 \Pi_{j=1}^3 d\rho_j \, f\biggl(\sum_{j=1}^3 \rho_j \biggr) \biggl\langle |{\hat \Delta}_p|^4 \, \biggl(\frac{w_y}{v_F} \biggr)^2 \biggl[ - \frac{3}{2} \biggl(\sum_{j=1}^3 \rho_j^2 \biggr) \Pi_{k=1}^3 {\rm cos}(k) + \frac{1}{2} \sum_{i \neq j} \biggl(3 \rho_i \rho_j {\rm sin}(i) {\rm sin}(j) {\rm cos}(6-i-j) \nonumber \\
&+& (-1)^{i+j} {\rm cos}(6-i-j) \biggl( \rho_i \rho_j \, {\rm cos}(i) {\rm cos}(j) - \frac{1}{2} (\rho_{6-i-j})^2 \, {\rm sin}(i) {\rm sin}(j) \biggr) \biggr) 
\biggr] \exp\biggl(-\frac{1}{2}\biggl(-\frac{1}{2}R_{24} + R_{14} \biggr) \biggr) \, {\rm cos}(I_4) \biggr\rangle_{\rm FS}, \nonumber \\
{\overline L}_3 &=& - \frac{1}{4} \int_0 \Pi_{j=1}^3 d\rho_j \, f\biggl(\sum_{j=1}^3 \rho_j \biggr) \, \rho_1 \, \biggl\langle |{\hat \Delta}_p|^4 \, \frac{w_x w_y}{v_F^2} \, \sum_{i \neq j} \biggl[ \rho_{6-i-j} \, {\rm sin}(6-i-j) ( \, 3 {\rm cos}(i) {\rm cos}(j) + (-1)^{i+j} {\rm sin}(i) {\rm sin}(j) \, ) \nonumber \\
&-& 2 (-1)^{i+j} {\rm cos}(6-i-j) \, \rho_i \, {\rm cos}(i) {\rm sin}(j) \biggr] \, \biggl( \, {\rm Re}(\mu) \, (3\rho_1-\rho_3+\rho_2) \, {\rm cos}(I_4) + {\rm Im}(\mu) \, \biggl( \sum_{j=1}^3 \rho_j \biggr) \, {\rm sin}(I_4) \biggr) \nonumber \\
&\times& \exp\biggl(-\frac{1}{2}\biggl(-\frac{1}{2}R_{24} + R_{14} \biggr) \biggr\rangle_{\rm FS}.
\label{tildeLj}
\end{eqnarray}
Actual numerical calculations of tilt moduli are performed according to eq.(\ref{elascon}) in terms of $L_j$ and ${\overline L}_j$ ($j=1$, $2$, and $3$) given above. 

Before ending this section, we point out a couple of essential features appearing in the coefficients of the GL free energy and the elastic constants derived above. First of all, noting that the contributions of the orbital depairing appear as $|\mu| \rho_n$ or ${\rm Re}(\mu) \rho_n$ everywhere, the {\it effective} strength of the paramagnetic depairing is $\alpha_{M} \, ( \, h \, \gamma \, )^{1/2}$. Hence, an increase of the flux density $B$ or of the anisotropy $\gamma$ enhances the paramagnetic depairing effects. Further, consistently with this discussion, the period of the FFLO modulation is scaled by not the coherence length but the magnetic length $r_B \gamma^{1/2}$ (see eq.(\ref{scaledQ})), implying that the period of the modulation decreases with increasing $B$. Although this is not surprising because the paramagnetic effect is enhanced with increasing $B$, one should note that $r_B$ does not arise in any approach in the Pauli limit with no vortices included. 

\section{III. Numerical results and discussions} 

In this section, typical numerical results following from the expressions of the coefficients, eq.(\ref{GLcoef2}), and the elastic constants, eq.(\ref{elascon}), are presented, and their relevance to an available experimental result \cite{ISSP} will be discussed. 

In the ensuing numerical results, the values $\alpha_M = 10.65$ and $\gamma=4.5$ were used for ${\bf H} \parallel {\hat x}$, otherwise stated. Further, for comparison, elastic constants in perpendicular fields, ${\bf H} \parallel {\hat z}$, were also examined in terms of the values $\alpha_M = 4.95$ and $\gamma=2.85$. It is straightforward to, with an appropriate replacement of FS and ${\bf w}$, obtain the corresponding expressions to eqs.(\ref{GLcoef2}) and (\ref{elascon}) in perpendicular fields which were used elsewhere \cite{RIM2S,IA}. The orbital-limiting field $H_{\rm orb}(0)$ in each field configuration was estimated numerically from $a_0(0)=0$ with $\alpha_M=0$ in $T \to 0$. 

%%%%%%%%%%%%%%%%%%%
\begin{figure}[b]
\scalebox{0.3}[0.3]{\includegraphics{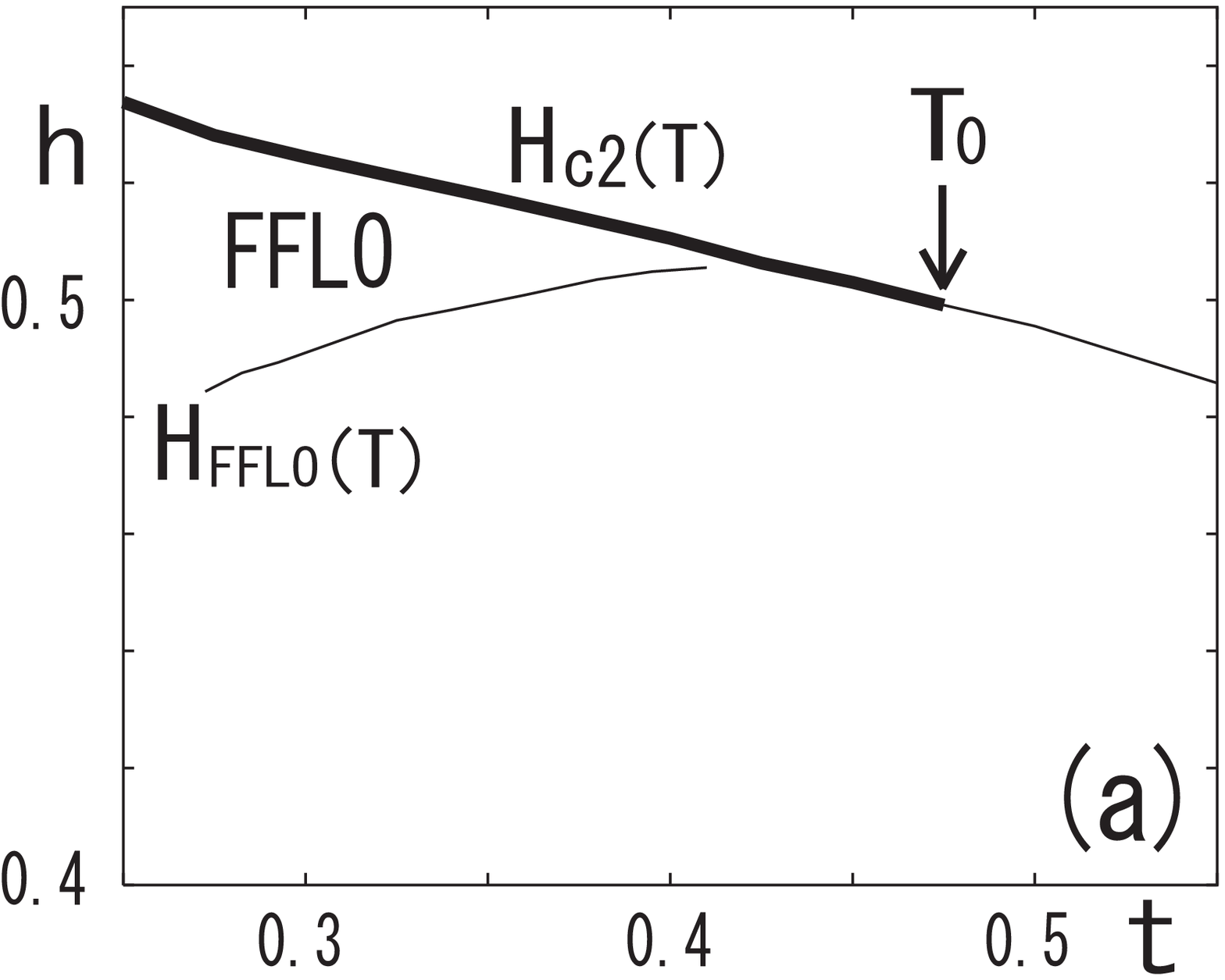}}
\scalebox{0.3}[0.3]{\includegraphics{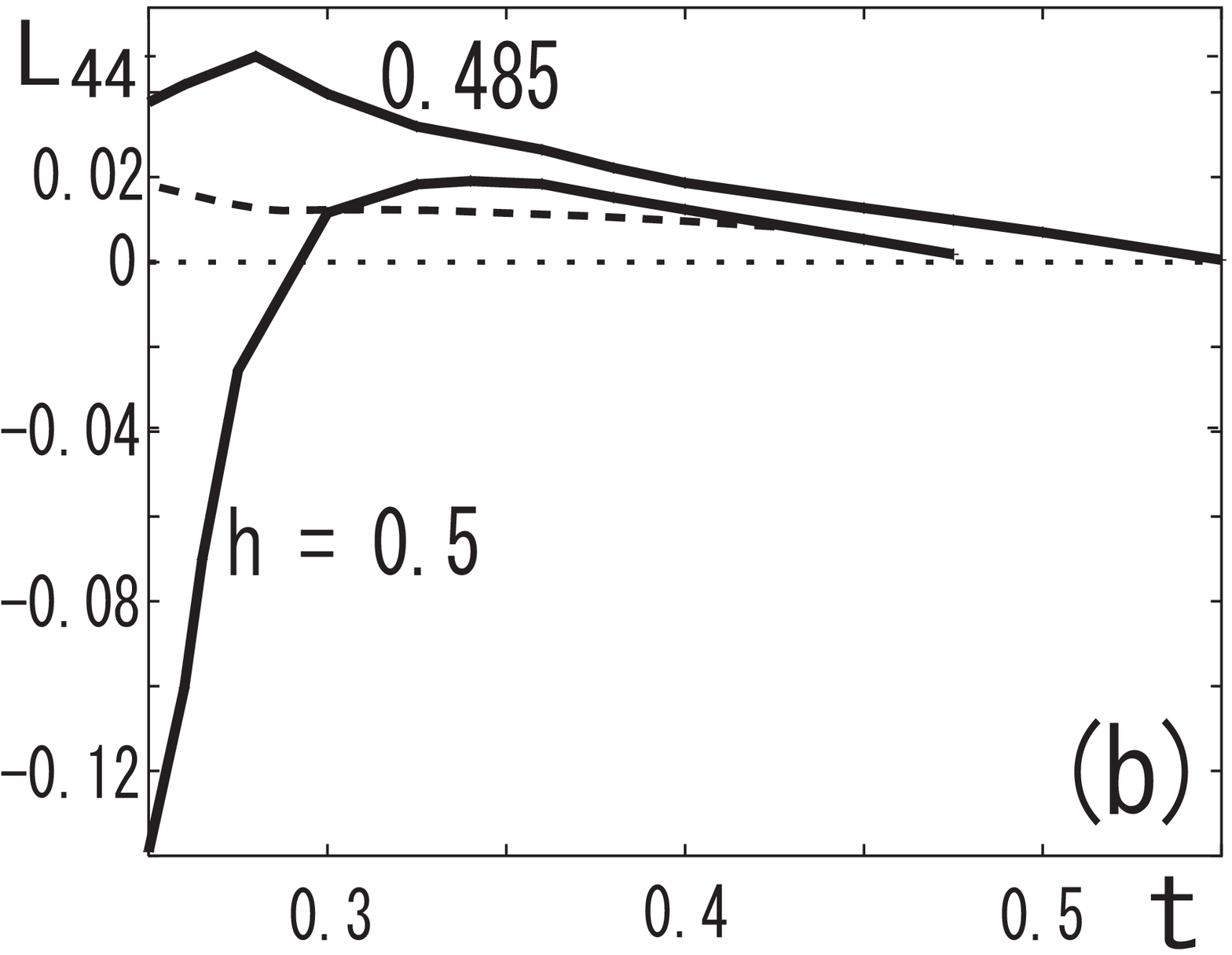}}
\caption{(a) Example of the $h$-$t$ mean field phase diagram in ${\bf H} \parallel {\hat x}$ obtained numerically, where each thick (thin) solid curve is the discontinuous (2nd order) mean field transition curve. The low temperature region in $t < 0.25$ where another FFLO-like vortex lattice \cite{RIM2S,RI} described by the $n=1$ LL modes of $\Delta_{\rm e}$ occurs is not shown here. (b) The corresponding numerical data of $L_{44}$, defined in the text, in clean limit. The lower (upper) solid curve denotes $L_{44}(t)$ in ${\bf H} \parallel 
{\hat x}$ for $h=0.5$ ($h=0.485$), while the dashed curve is that in ${\bf H} \parallel {\hat z}$ given in Fig.2 and follows from the dotted curve in Fig.4. 
 } \label{fig.3}
\end{figure}

In determining the phase diagram Fig.3(a), the onset temperature $T_0$ of the discontinuous $H_{c2}$-transition was determined as the position at which $V_4(q)$ changes the sign while verifying $V_6(q) > 0$, and $H_{\rm FFLO}(T)$ is determined, when $c^{(4)}(\alpha_{\rm e}) > 0$, as the line on which $c^{(2)}(\alpha_{\rm e})$ in $H < H_{c2}$ changes the sign. The above-mentioned conditions on the sign of $V_6$ and $c^{(4)}$ were satisfied in all of the resulting numerical data in ${\bf H} \parallel {\hat x}$. 
As reported elsewhere \cite{IA}, the second order transition on $H_{\rm FFLO}(T)$ occurs at lower temperatures than $T_0$ and {\it decreases} upon cooling, since, as mentioned at the end of sec.II, the paramagnetic depairing is stronger in higher fields and at lower temperatures. We note that, in the case of Fig.4 (a), the FFLO state modulating along ${\bf H}$ is overcome, in $t < 0.25$, by another FFLO-like vortex state with a modulation {\it perpendicular} to ${\bf H}$ and formed in the next ($n=1$) LL \cite{RIM2S,RI}. However, this another FFLO-like state has no periodic modulation parallel to ${\bf H}$, implying that no specific feature is expected in tilt deformations. For this reason, we focus here on the higher temperature range in which the $n=1$ LL state does not occur. 

%%%%%%%%%%%%%%%%%%%%%%%%%%%%%%
\begin{figure}[t]
\scalebox{0.3}[0.3]{\includegraphics{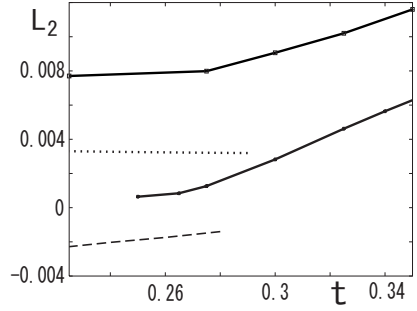}}
\caption{Numerical results of $L_2$ in ${\bf H} \parallel {\hat x}$ for $h=0.485$ (upper solid curve) and $h=0.5$ (lower solid one) and in ${\bf H} \parallel {\hat z}$ for $h=0.24$ (dotted one) obtained by including contributions from the GL-{\it quartic} term to the description of the FFLO state. The dashed line showing $L_2 < 0$ follows from eq.(\ref{Lj}) with $\gamma=3.65$.} \label{fig.4}
\end{figure}
%%%%%%%%%%%%%%%%%

The main result in this paper is seen in Fig.3 (b): 
As the curve of $L_{44}(T)$ in $h=0.5$ shows, where 
\begin{equation}
L_{44} \equiv = \frac{r_B^2 \gamma}{2 \pi^2 \xi_0^2} \biggl( E_1 - 4\frac{E_3^2}{E_2} \biggr),
\label{L44}
\end{equation} 
$\delta C_{44}$ proportional to $L_{44}$ (see eq.(\ref{delc44})) can become {\it negative} in the FFLO state and thus, leads to a reduction of $C_{44}$. As argued in sec.I, this $\delta C_{44}$-reduction is more remarkable in higher $B$ and at lower $T$, i.e., as the paramagnetic depairing is stronger. 

In Fig.4, results of the nodal plane's tilt modulus $E_2$ in various 
situations are shown. When ${\overline L}_2$ is neglected in $E_2$, and only the contribution $L_2$ from ${\cal F}_2$ is kept, we often see, as in the dashed curve, negative $E_2$ values. It implies that theoretical approaches based on an evaluation of an inhomogenuity of the order parameter field $\Delta({\bf r})$ only from the quadratic GL term ${\cal F}_2$ \cite{HM} cannot describe the stability of the FFLO state properly as far as the normal to FFLO transition is discontinuous in the mean field approximation. In contrast, once ${\overline L}_2$ is included in $E_2$, the results of $E_2$ we obtain remain positive, as is physically required for a stable FFLO state modulating along ${\bf H}$, although they significantly decrease upon cooling especially in higher fields, reflecting the "softening" of nodal planes induced by the paramagnetic depairing (see sec.I). Note that the decrease of $E_2$ is not unlimited according to our microscopic calculation, suggesting that a genuine instability of the FFLO state modulating along ${\bf H}$ does not necessarily occur. In any case, the primary origin of the negative $\delta C_{44}$ in the FFLO state is clearly this reduction of $E_2$ upon cooling which becomes more remarkable in situations affected by a stronger paramagnetic depairing realizable in ${\bf H} \parallel {\hat x}$. In this way, the physical consideration given in sec.I is supported by the microscopic derivation of the elastic moduli in sec.II. 

The reduction of $C_{44}$ is estimated based on 
\begin{eqnarray}
\delta C_{44} = \frac{B^2}{4 \pi} \, R(B, T) \frac{r_B^2}{(\lambda(0))^2}, 
\label{scalec44}
\end{eqnarray}
where $\lambda(0)$ is the London penetration depth at $T=0$, and 
\begin{eqnarray}
R(B, T) = \biggl(E_1 - 4 \frac{E_3^2}{E_2} \biggr) \cdot \biggl[ \frac{1 - T/T_c}{2 E_1(B \to 0; T \to T_c)} 
\biggr]. 
\label{RHT}
\end{eqnarray}
The expression in the bracket $[ \,\,\, ]$ is evaluated as 3.5 in terms of the $L_1$-result known within the weak-coupling BCS model. Based on Fig.3 (b), $R$ can reach $- \, 0.5$ in the temperature range just below $H_{\rm FFLO}(T)$. Imagining CeCoIn5 in several tesla and, as a rough estimation, taking the values $\lambda(0) \simeq 10^3$ (A) and $r_B \simeq 10^2$ (A), the resulting $- \delta C_{44}$ values in the FFLO state near $H_{\rm FFLO}(T)$ are two order of magnitude smaller than $B^2/(4 \pi)$. This small value of reduction of $\delta C_{44}/C_{44}$ will partly become larger by taking account of the dispersive main term \cite{IR92} of $C_{44}$ (the ${\bf k}$ dependences of the $B^2/4\pi$ term). 

  The present result may be relevant to the observation in the ultrasound measurements \cite{ISSP}, where the normalized sound velocity $v_s$ has shown a reduction upon cooling through $H_{\rm FFLO}$ for the displacements of the underlying crystal perpendicular to ${\bf H}$ (Lorentz mode), while no signature of a comparable magnitude has been seen for displacements parallel to ${\bf H}$ (nonLorentz mode). This remarkable anisotropy of phenomena is an evidence of a structural change on $H_{\rm FFLO}(T)$ of the vortex state. Since, strictly speaking, a pinning of the nodal planes due to the crystal lattice is present in the latter, this fact implies that the pinning of the nodal planes is quantitatively negligible. In contrast, in the Lorentz mode, not the nodal planes but the vortices couple to the crystal displacement, and consequently, the observed reduction of $v_s$ in this mode in entering the FFLO state implies some reduction of $C_{44}$ and/or of the vortex pinning strength. First, the {\it overall} temperature variation of $v_s$ surviving at low enough $t$ likely reflects that of the order parameter amplitude $\alpha_{\rm e}$ carried by the pinning strength (see Fig.3 in Ref.\cite{ISSP}). In addition, the data of quantities measuring $\alpha_{\rm e}(T)$ \cite{Capan,Martin,Brazil,comcapan} show a reduction of $\alpha_{\rm e}(T)$, due to an increase of $q$ \cite{RI}, upon entering the FFLO state compared to the extrapolation of $\alpha_{\rm e}$ in the Abrikosov state to lower temperatures. Thus, the {\it relative} reduction of $\alpha_{\rm e}$ and hence of the pinning strength may be one origin of the $v_s$-reduction in entering the FFLO state. On the other hand, the reduction of $C_{44}$, obtained in the present work, upon entering the FFLO state, also leads to a reduction of $v_s$ in the Lorentz mode \cite{ISSP}. The feature seen in Fig.3 that the $C_{44}$-reduction is more remarkable in higher fields is consistent with the field dependence of the observed 
anomalies \cite{ISSP}. Although the reduction of $C_{44}$ estimated above seems to be too small to quantitatively explain the observed reduction of $v_s$, we expect this discrepancy to be partly resolved by going beyond the GL expansion in $\Delta$ because $\alpha_{\rm e}^2$ has been presumably underestimated in the GL analysis of the present case with a discontinuous $H_{c2}$-transition. 

For comparison, $\delta C_{44}$ in ${\bf H} \parallel {\hat z}$ was also examined by adding a small {\it noncylindrical} portion with large $|w_z/w_x|$, stabilizing \cite{RIM2S,RI} an FFLO in ${\bf H} \parallel {\hat z}$, to the cylindrical FS. Although, as given in Fig.2, it shows the {\it familiar} type of softening behavior of $\delta C_{44}$ at $H_{\rm FFLO}$, this cusp-like feature is not visible on the scale of Fig.4 (see the dotted curve), and, as shown there, the temperature variations of $\delta C_{44}$ and $L_2$ are quite weak around $H_{\rm FFLO}$. Thus, changes of the tilt response through $H_{\rm FFLO}$ in ${\bf H} \parallel {\hat z}$ are negligible. The presence in ${\bf H} \parallel {\hat x}$ and the absence in ${\bf H} \parallel {\hat z}$ of a clear peak effect in magnetization data of CeCoIn5 near $H_{\rm FFLO}$ \cite{Brazil} might be related to the corresponding difference in $C_{44}$ mentioned above, because a peak effect occurs reflecting a notable change of a vortex elastic modulus \cite{LO}.  

Throughout this paper, we have focused on the case in which vortex lattices {\it in equilibrium} are described in the lowest LL. As shown elsewhere, \cite{RIM2S,RI} a higher LL vortex lattice tends to occur in "clean limit", which is defined as the case with infinitely long quasiparticle's (QP's) mean free path, and in the case with large $\alpha_M$. In fact, in Fig.3 (a), the $n=1$ LL vortex lattice with additional nodal lines in the plane perpendicular to ${\bf H}$ occurs in much lower temperatures than the range shown there. Such an appearance of higher LL states due to a large $\alpha_M$ under a {\it strictly parallel} field to the layers is closely related to a different issue \cite{2dpauli} of transitions between 2D vortex lattices in the large $\gamma$ limit induced by a tilt of the applied field from the parallel field configuration. However, the $n=1$ LL state in the present case is easily pushed down to $T=0$ due to a finite but nevertheless quite long QP's mean free path and is expected not to occur at measurable temperatures in CeCoIn5 \cite{RI}. In relation to this, it will be valuable to point out that the 2D higher LL vortex lattices due to the tilt of the applied field \cite{2dpauli} may be expected only when the opposite relation to eq.(\ref{paulidame}), i.e., 
\begin{equation}
\gamma \gg \frac{1.3 \xi_0^2}{d^2} \alpha_M^{({\rm 2D})}, 
\label{paulidame2}
\end{equation}
is satisfied. In fact, the neglect \cite{2dpauli} of the orbital depairing effect in the case with a strictly parallel field is equivalent to assuming the absence of structural transitions between different Josephson vortex lattices, and this assumption is justified only in higher fields than $1/(2e \gamma d^2)$ \cite{II}. For CeCoIn5 with a weak anisotropy \cite{IA2}, eq.(\ref{paulidame2}) is never satisfied. 

In conclusion, the vortex tilt modulus $C_{44}$ in the FFLO state modulating along ${\bf H}$ may be reduced due to tilts of the nodal planes. This $C_{44}$-reduction should be remarkable especially in quasi 2D materials with a strong paramagnetic depairing in the parallel field, in which the nodal planes are fixed only weakly by the field direction, and may be an origin of the reduction of sound velocity in the Lorentz mode upon entering the FFLO phase observed in CeCoIn5.

\begin{acknowledgements}
The author is grateful to Y. Matsuda and C. J. van der Beek for discussions on this issue. 
\end{acknowledgements}

\appendix

\section{Appendix A} 

Strictly speaking, $a_0(0)$ and other coefficients in eq.(\ref{GLcoef2}), described within the $n=0$ LL, are affected by a mixing of the $n=2$ LL modes in expressing the equilibrium order parameter $\Delta_e$. For instance, $a_0(0)$ should be replaced by $a_0(0) - (a_{20})^2/a_2(0)$, where 
\begin{eqnarray}
a_2(0) &=& \frac{1}{2}{\rm ln}(h) + \int_0^\infty d\rho \biggl[ \frac{1}{\rho} e^{-(\pi \xi_0 \rho/r_B)^2} \nonumber \\
&-& \, f(\rho) \biggl\langle |{\hat \Delta}_p|^2 \biggl( 1 - 2 \rho^2 |\mu|^2 + \frac{\rho^4 |\mu|^4}{2} \biggr) \exp\biggl(-\frac{|\mu|^2 \rho^2}{2} \biggr) \biggr\rangle_{\rm FS} \biggr], \nonumber \\
a_{20} &=& \int_0^\infty d\rho \, f(\rho) \biggl\langle |{\hat \Delta}_p|^2 \rho^2 \mu^2 \, \exp\biggl(-\frac{|\mu|^2 \rho^2}{2} \biggr) 
\biggr\rangle_{\rm FS}. 
\end{eqnarray}

\section{Appendix B}

We work here in the conventional GL free energy 
\begin{eqnarray}
{\cal F}_{\rm GL} = \int d^3r \biggl[ \frac{T-T_c}{T_c} |\Psi|^2 
+ \xi_0^2 \, \biggl[ \sum_{j=x, y} |(-{\rm i}\partial_j 
+ 2e {\bf A}_j)\Psi|^2 + 
\gamma^{-2} |(-{\rm i}\partial_z + 2e A_z) \Psi|^2 \, \biggr] + \frac{b}{2}|\Psi|^4 + \frac{1}{8 \pi} ({\rm curl} \delta {\bf A})^2 \biggr],
\label{b1}
\end{eqnarray}
where $b > 0$, and ${\bf A} = B y {\hat z} + \delta {\bf A}(x)$. Here, the order parameter $\Psi$ is assumed to be the sum of the Abrikosov solution $\Psi_{\rm A} = \alpha_e \Psi^{(0)}$ in the $n=0$ LL and its excitation $\alpha_e \delta a_1(x) \Psi^{(1)}$ in the $n=1$ LL, where $\Psi^{(n)}=\Pi_+^n \Psi^{(0)}/\sqrt{n!}$. Noting that $(\Pi_+ \pm \Pi_-) \Psi = \alpha_e (\sqrt{2} \delta a_1 \Psi^{(2)}+\Psi^{(1)} \pm \delta a_1 \Psi^{(0)})$ and assuming the normalization $\int_{\bf r} (\Psi^{(n)})^* \Psi^{(m)} = \delta_{n,m}$, the gradient terms dependent on $\delta a_1$ in eq.(\ref{b1}) are rewritten in the form 
\begin{eqnarray}
{\cal F}_{\rm GL}|_{\rm grad} &=&  \alpha_e^2 \xi_0^2 \gamma^{-1} \int dx \biggl( \, \biggl[ \biggl( \frac{2e}{\gamma^{1/2}} \delta A_z(x) + \frac{\sqrt{2}}{r_B} {\rm Re}\delta a_1(x) \biggr)^2 + \biggl(2e \gamma^{1/2} \delta A_y(x) - \frac{\sqrt{2}}{r_B} {\rm Im}\delta a_1(x) \biggr)^2 \biggr] \nonumber \\
&+& \gamma |\partial_x \delta a_1|^2 + r_B^{-2} |\delta a_1|^2 \biggr). 
\label{b2}
\end{eqnarray}
In obtaining the $|\partial_x \delta a_1|^2$ term, the fact that, since $\int_{\bf r} (\Psi^{(0)})^* \Psi^{(1)} =0$, $\delta a_1$ decouples with $\delta A_x$ up to the harmonic order was used. The last term can be absorbed into the first term of eq.(\ref{b1}) so that the mass term vanishes not at $T_c$ but on the straight $H_{c2}(T)$ line. As shown in Ref.\cite{IR92}, no quadratic terms in $\delta a_1$ occur from the sum of the resulting $|\psi|^2$ term and $b|\psi|^4/2$. 

Here, let us first examine eq.(\ref{b2}) by neglecting the $x$ dependences. Then, when 
\begin{equation}
\delta a_1 = \frac{ \gamma^{-1/2} u + {\rm i} \gamma^{1/2} v}
{\sqrt{2} \, r_B}, 
\label{b3}
\end{equation}
where ${\bf u} \equiv u {\hat y} + v {\hat z}$ is the vortex displacement field, the famous Josephson relation 
\begin{equation}
\delta {\bf A} = {\bf u} \times {\bf B}, 
\label{b4}
\end{equation}
or ${\bf E} = - (\partial {\bf u}/\partial t) \times {\bf B}$ 
implying a nonvanishing vortex flow resistance, follows as a condition for minimizing the fluctuation free energy. In fact, this Josephson relation implies the vanishing of the static superfluid rigidity $\Upsilon_{s \, \perp}$ to a current perpendicular to ${\bf H}$, because this relation implies that $\delta {\bf A}$ is lost (eaten) by the $n=1$ LL fluctuation from ${\cal F}_{\rm GL}|_{\rm grad}$ so that $\Upsilon_{s \, \perp} \propto \delta^2 {\cal F}/\delta (\delta {\bf A})^2 =0$. 

Next, the $x$ dependences of the fluctuation fields will be incorporated. 
Further, by substituting eq.(\ref{b4}) into the magnetic energy term (the last term of eq.(\ref{b1})), the main term $B^2/(4 \pi)$ of the vortex tilt modulus $C_{44}$ is obtained from there, while $\delta C_{44}$ defined in sec.I follows from the remaining term in eq.(\ref{b2}), i.e., 
\begin{equation}
\alpha_e^2 \xi_0^2 |\partial_x \delta a_1|^2 = \frac{\xi_0^2}{2 r_B^2 \gamma} \alpha_e^2 (\partial_x u)^2 = \frac{B}{32 \pi e \gamma (\lambda(T))^2} \biggl(1 - \frac{B}{H_{c2}(T)} \biggr) \, (\partial_x u)^2, 
\label{b5}
\end{equation}
where $\lambda(T)$ is the London penetration depth. It is easy to see that the resulting expression 
\begin{equation}
\delta C_{44} = \frac{B}{16 \pi e \gamma (\lambda(T))^2} \biggl(1 - \frac{B}{H_{c2}(T)} \biggr)
\label{b6}
\end{equation}
is $- BM$, where $M$ is the magnetization following from the SC condensate. Then, the familiar total tilt modulus 
\begin{equation}
\frac{B(B-4 \pi M)}{4 \pi} = \frac{BH}{4 \pi} 
\end{equation}
is obtained for this Abrikosov vortex lattice. In general, $\delta C_{44}$ does not have to be equivalent to $-MB$: As shown in the text, $\delta C_{44}$ can become negative in the FFLO vortex lattice, while $-M = \partial {\cal F}_e \, /\partial B$ of the FFLO vortex lattice is always positive because 
the magnetic 
field tends to destroy superconductivity and hence, increases the condensation energy.


\section{References}

\begin{thebibliography}{99}

\bibitem{Bianchi} A. Bianchi, R.Movshovich, C.Capan, P.G.Pagliuso, and J.L.Sarrao, Phys. Rev. Lett. {\bf 91}, 187004 (2003). 
\bibitem{Capan} C. Capan, A. Bianchi, R. Movshovich, A. D. Christianson, A. Malinowski, M. F. Hundley, A. Lacerda, P. G. Pagliuso, and J. L. Sarrao, Phys. Rev. B {\bf 70}, 134513 (2004). 
\bibitem{Martin} C. Martin, C. C. Agosta, S. W. Tozer, H. A. Radovan, E. C. Palm, T. P. Murphy, and J. L. Sarrao,  Phys. Rev. B {\bf 71}, 020503(R) (2005). 
\bibitem{Kaku} K. Kakuyanagi, M. Saitoh, K. Kumagai, S. Takashima, M. Nohara, H. Takagi, and Y. Matsuda Phys. Rev. Lett. {\bf 94}, 047602 (2005). 
\bibitem{AI} H. Adachi and R. Ikeda, Phys. Rev. B {\bf 68}, 184510 (2003). 
\bibitem{IA2} R. Ikeda and H. Adachi, Phys. Rev. Lett. {\bf 95}, 269703 
(2005). 
\bibitem{Pauli} P. Fulde and R. A. Ferrell, Phys. Rev. {\bf 135}, A550 (1964); A. I. Larkin and Y. N. Ovchinnikov, Sov. Phys. JETP {\bf 20}, 762 (1965). 
\bibitem{RIM2S} R. Ikeda, cond-mat/0610863 (for the Proceedings of M$^2$S, Dresden. To appear in Physica C). 
\bibitem{Izawa} K. Izawa, H.Yamaguchi, Y.Matsuda, H.Shishido, R.Settai, and 
Y.Onuki, Phys Rev. Lett. {\bf 87}, 057002 (2001). 
\bibitem{ISSP} T. Watanabe, Y. Kasahara, K. Izawa, T. Sakakibara, Y. Matsuda, C. J. van der Beek, T. Hanaguri, H. Shishido, R. Settai, and Y. Onuki, 
Phys. Rev. B {\bf 70}, 020506(R) (2004). 
\bibitem{com2}  This assumption is reasonable in ultrasound measurements with a long enough wavelength of the sound. 
\bibitem{comhb} Except in the analysis on the vortex tilt modulus in Appendix B, the magnitude $H$ of the external field ${\bf H}$ can be identified in this paper 
with the uniform flux density $B$. 
\bibitem{II} R.Ikeda and K.Isotani, J. Phys. Soc. Jpn. {\bf 67}, 983 (1998) ; R. Ikeda, J. Phys. Soc. Jpn. {\bf 71}, 587 (2002).  
\bibitem{comUji} When the vortex lattice with a modulation induced by the paramagnetic depairing has, in contrast to the case of CeCoIn$_5$, nodal planes {\it parallel} to ${\bf H}$, the structural transitions between Josephson vortex lattices play important roles even if this inequality is satisfied. This case is relevant to the corresponding issue in $\lambda$-(BET)$_2$FeCl$_4$ [S. Uji, T. Terashima, M. Nishimura, Y. Takahide, T. Konoike, K. Enomoto, H. Cui, H. Kobayashi, A. Kobayashi, H. Tanaka, M. Tokumoto, E. S. Choi, T. Tokumoto, D. Graf, and J. S. Brooks, Phys. Rev. Lett. {\bf 97}, 157001 (2006)] and will be discussed in a separate paper. 
\bibitem{PAL} P.A. Lee and M.G. Payne, Phys. Rev. B {\bf 5}, 923 (1972). 
\bibitem{IA} R.Ikeda and H.Adachi, Phys. Rev. B {\bf 69}, 212506 (2004). 
\bibitem{IR92} R. Ikeda, T. Ohmi, and T. Tsuneto, J. Phys. Soc. Jpn. {\bf 61}, 254 (1992). See also R. Ikeda, J. Phys. Soc. Jpn. {\bf 64}, 3825 (1995). 
\bibitem{RI} R. Ikeda, cond-mat/07060321. 
\bibitem{HM} M. Houzet and V. P. Mineev, cond-mat/0703104. 
\bibitem{Brazil} X. Gratens, L. Mendonca Ferreira, Y. Kopelevich, N. F. Oliveira Jr., P. G. Pagliuso, R. Movshovich, R. R. Urbano, J. L. Sarrao, J. D. Thompson, cond-mat/0608722. 
\bibitem{comcapan} We stress that a kink anomaly \cite{Capan} at $H_{\rm FFLO}$ in the thermal conductivity for a heat current $\parallel {\bf H}$ should appear even when the nodal planes are perpendicular to the vortex lines because, in GL region, the thermal conductivity usually increases as, just like in entering the FFLO state by increasing the field \cite{RI}, the gap amplitude is reduced. See, for instance, the work by C. Caroli and M. Cyrot [ Phys. Kondensierten Materie {\bf 4}, 285 (1965) ]. Thus, the observation in Ref.2 does not contradict the picture with the FFLO state modulating along the field. 
\bibitem{LO} A. I. Larkin and and Y. N. Ovchinnikov, Sov. Phys. JETP {\bf 34}, 409 (1979). 
\bibitem{Maki} H. Won, K. Maki, S. Haas, N. Oeschler, F. Weickert, and P. Gegenwart, Phys. Rev. B {\bf 69}, 180504(R) (2004). 
\bibitem{2dpauli} See, for instance, U. Klein, D. Rainer, and H. Shimahara, J. Low Temp. Phys. {\bf 118}, 91 (2000); M. Houzet, A. Buzdin, L. Bulaevskii, and M. Maley, Phys. Rev. Lett. {\bf 88}, 227001 (2002); U. Klein, Phys. Rev. B {\bf 69}, 134518 (2004); K. Yang and A. H. MacDonald, Phys. Rev. B {\bf 70}, 094512 
(2004). 

\end{thebibliography}
\end{document}